\documentclass{DISproc}

\def\beq{\begin{equation}}
\def\eeq{\end{equation}}
\def\beqa{\begin{eqnarray}}
\def\eeqa{\end{eqnarray}}

\begin{document}
%------------------------------------
\title{Differential and total cross sections for top pair and single top production}

%for single authors the superscripts are optional
\author{{\slshape Nikolaos Kidonakis}\\[1ex]
Kennesaw State University, Physics \#1202, Kennesaw, GA 30144, USA}

% please enter the contribution ID for the DOI
\contribID{251}

\doi  % if there is an online version we will register DOIs

\maketitle

\begin{abstract}
I present theoretical results at approximate NNLO from NNLL resummation 
for top quark production 
at the LHC and the Tevatron, including new results at 8 TeV LHC energy. 
Total cross sections are shown for $t{\bar t}$ production, for single top 
production in the $t$ and $s$ channels and via associated $tW$ production, 
and for associated $tH^-$ production. 
Top quark transverse momentum and rapidity distributions in 
$t{\bar t}$ production are also presented, as well as new results for 
$t$-channel single top and single antitop $p_T$ distributions.
\end{abstract}

\section{Introduction}

The top quark is a centerpiece of LHC and Tevatron physics, and 
both $t{\bar t}$ and single top production are being studied.
The LO partonic processes for top-antitop pair production are 
$q{\bar q} \rightarrow t {\bar t}$, dominant at the Tevatron, 
and $gg \rightarrow t {\bar t}$, dominant at LHC energies.
For single top quark production the partonic channels are the  
$t$ channel: $qb \rightarrow q' t$ and ${\bar q} b \rightarrow {\bar q}' t$;
the $s$ channel: $q{\bar q}' \rightarrow {\bar b} t$; and 
associated $tW$ production: $bg \rightarrow tW^-$. A related process is 
the associated production of a top quark with a charged Higgs, 
$bg \rightarrow tH^-$.

Higher-order QCD corrections are significant for top quark production.
Soft-gluon emission corrections are dominant and have been resummed through
NNLL accuracy \cite{ttbar,singletop}. 
Approximate NNLO (and even higher-order \cite{NNNLO}) 
differential cross sections have been derived from the expansion of the 
NNLL resummed result for $t{\bar t}$ \cite{ttbar} and single top 
\cite{singletop} production.

There are several different approaches to resummation (for a detailed review 
see \cite{review}; yet another approach appeared later in \cite{Beneke}). 
The approach used here is the only NNLL calculation at the differential 
cross-section level using the standard moment-space resummation in pQCD.

We note that the threshold approximation works very well not only for Tevatron 
but also for LHC energies because partonic threshold is still important.
There is less than 1\% difference between NLO approximate and exact cross 
sections, and this is also true for differential distributions, see 
the left plot in Fig. 1.

\section{$t{\bar t}$ production}

We begin with top-antitop pair production \cite{ttbar}. 
The NNLO approximate $t{\bar t}$ cross section at the Tevatron,  
with a top quark mass $m_t=173$ GeV and using MSTW2008 NNLO pdf \cite{MSTW}, is 
$7.08 {}^{+0.00}_{-0.24} {}^{+0.36}_{-0.27}$ pb, 
where the first uncertainty is from scale variation $m_t /2 <\mu <2m_t$ and 
the second is from the pdf at 90\% C.L. 

The $t{\bar t}$ cross section at the LHC at 7 TeV energy is 
$163 {}^{+7}_{-5} \pm 9$ pb; 
at 14 TeV it is $920 {}^{+50}_{-39}{}^{+33}_{-35}$ pb.  
The new result for the current 8 TeV LHC energy is 
\beqa
\sigma^{\rm NNLOapprox}_{t{\bar t}}(m_t=173\, {\rm GeV}, \, 8\, {\rm TeV})&=&234 {}^{+10}_{-7} \pm 12 \; {\rm pb} \, .
\nonumber 
\eeqa

\begin{figure}[htb]
  \centering
  \includegraphics[width=0.38\textwidth]{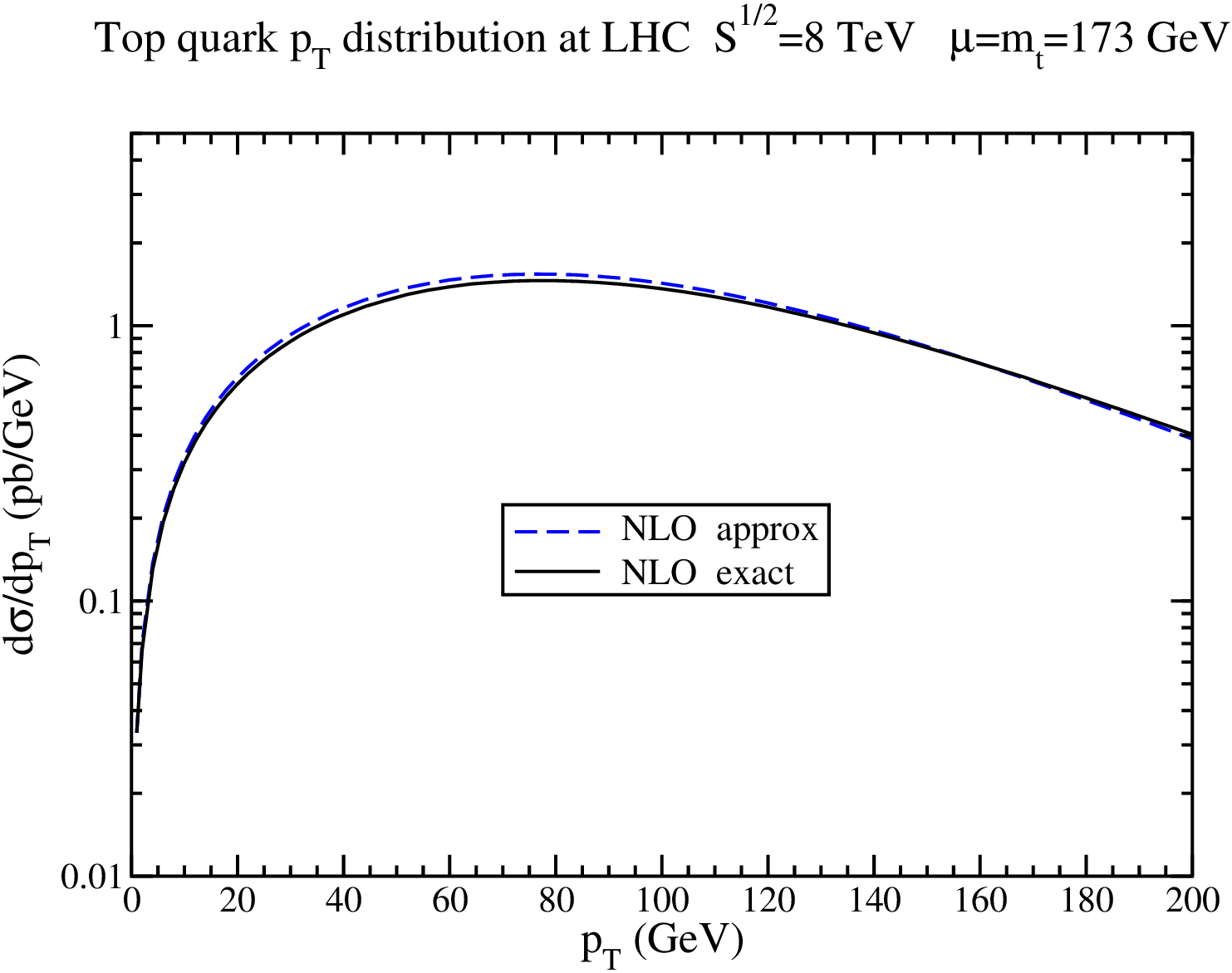}
\hspace{10mm}
  \includegraphics[width=0.38\textwidth]{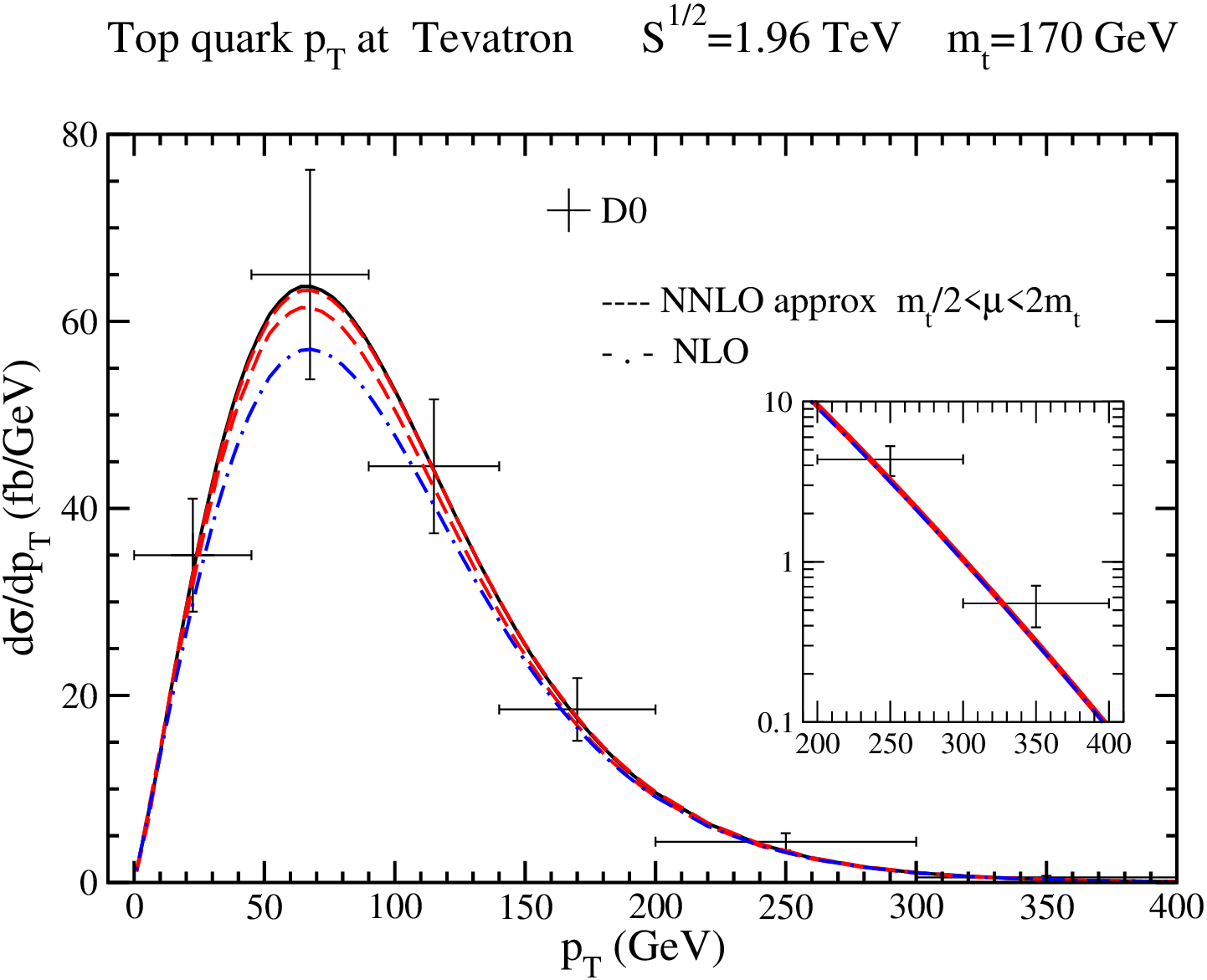}
  \caption{Top quark $p_T$ distribution at the LHC (left) and the Tevatron (right).}
  \label{toppTtev}
\end{figure}

The top quark $p_T$ distribution at the Tevatron is shown in the right plot 
of Fig. 1.
We note the excellent agreement of the NNLO approximate results with D0 data \cite{D0}.
The top quark $p_T$ distributions at the LHC at 7 and 8 TeV energies are 
shown in the left plot of Fig. 2.

\begin{figure}[htb]
  \centering
  \includegraphics[width=0.38\textwidth]{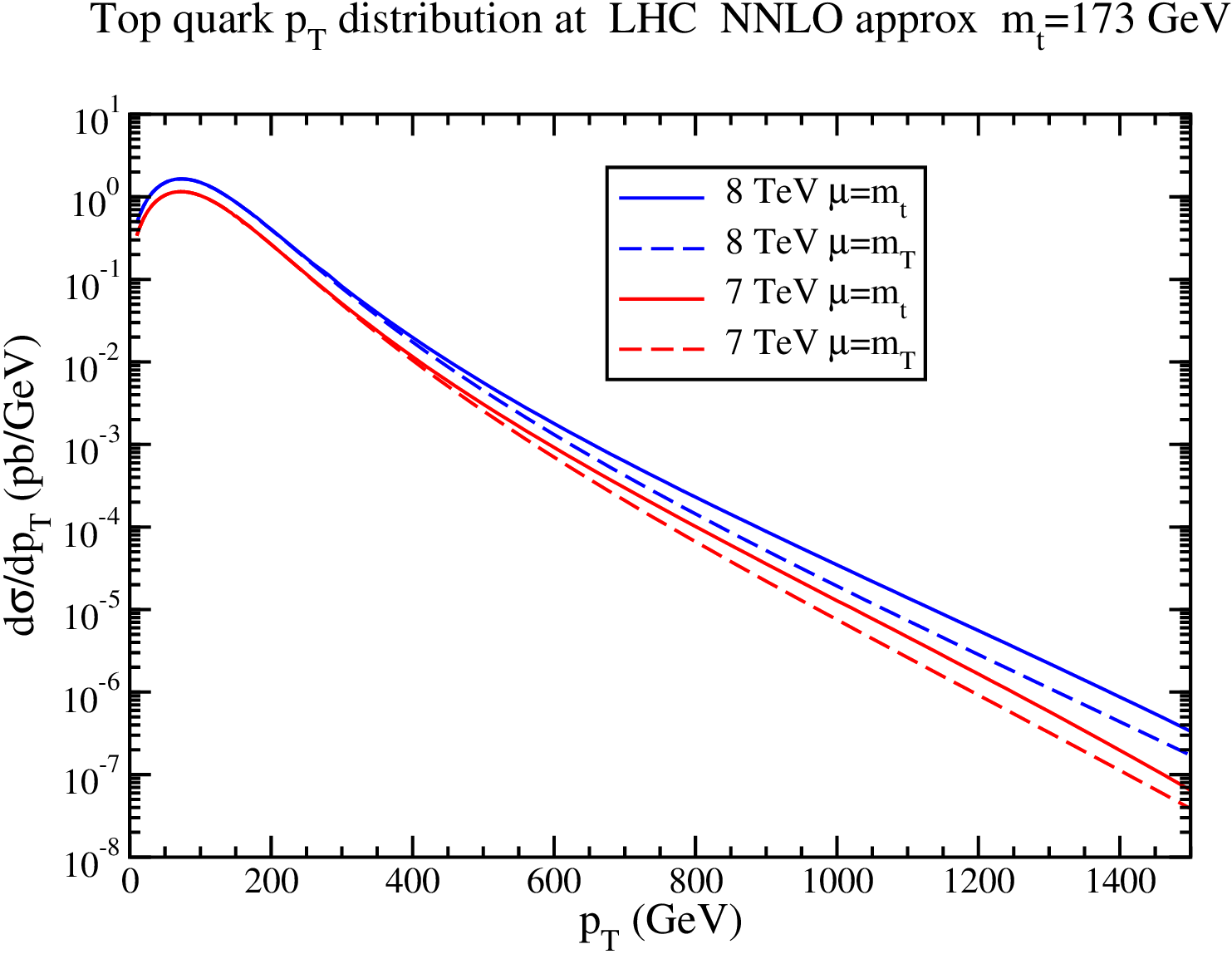}
\hspace{10mm}
  \includegraphics[width=0.38\textwidth]{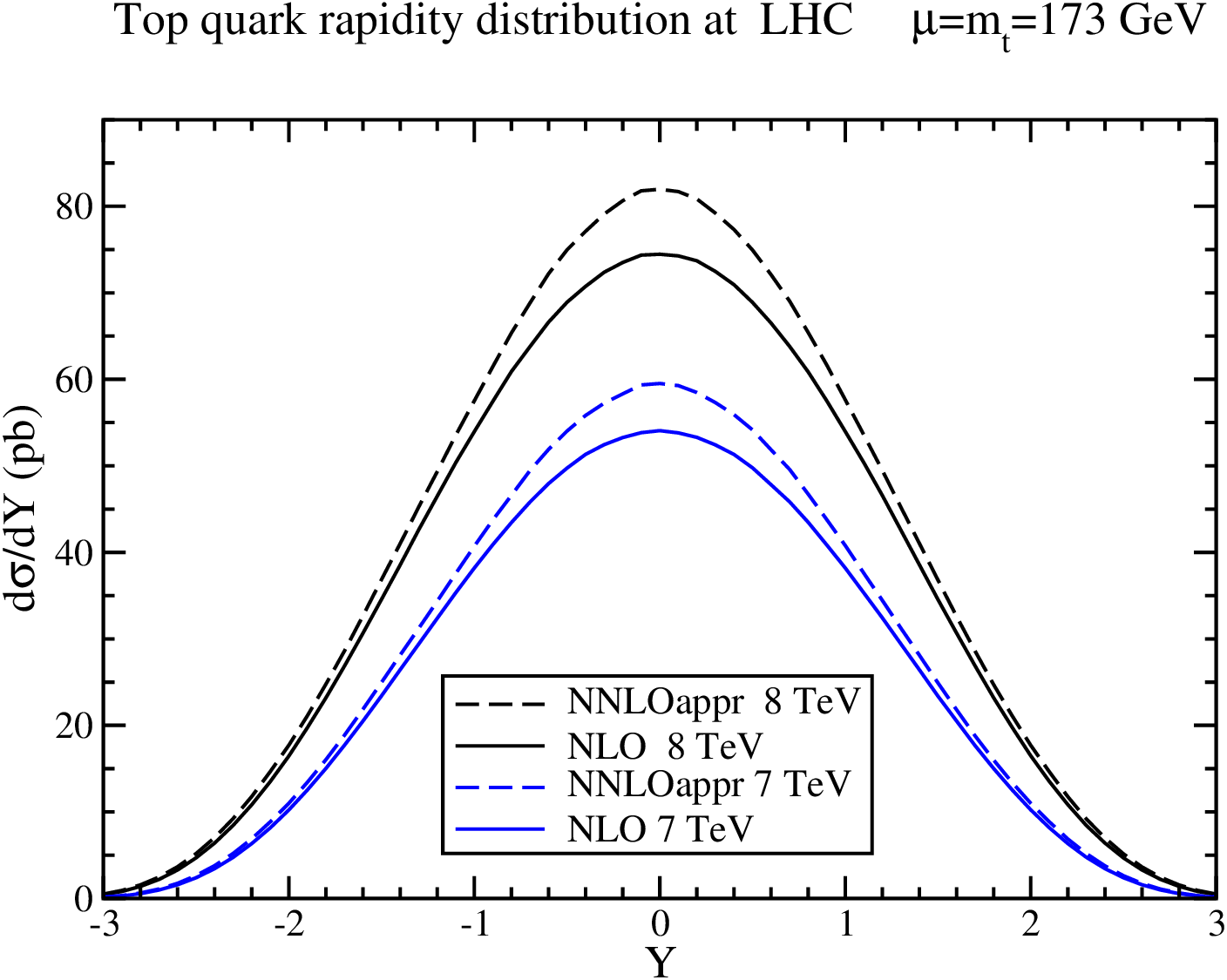}
  \caption{Top quark $p_T$ (left) and rapidity (right) distributions at the LHC  [$m_T=\sqrt{p_T^2+m_t^2}$].}
  \label{topylhc}
\end{figure}

The top quark rapidity distribution at the LHC at 7 and 8 TeV energies 
is shown in the right plot of Fig. 2.
The top quark rapidity distribution at the Tevatron displays a 
significant forward-backward asymmetry, 
$A_{\rm FB}=[\sigma(Y>0)-\sigma(Y<0)]/[\sigma(Y>0)+\sigma(Y<0)]=0.052^{+0.000}_{-0.006}$, which is smaller than observed values (see also the review in \cite{review} and recently \cite{BX,SWW}).

\section{Single top quark production}

We continue with single top production \cite{singletop}, 
and start with updated results for the $t$ channel.

\begin{table}
  \centering
\begin{tabular}{c|c|c|c}
\toprule 
{\scriptsize $t$-channel LHC}  & {\scriptsize $t$} &  
{\scriptsize ${\bar t}$} & {\scriptsize Total} \\ 
\midrule
{\scriptsize 7 TeV}  & {\scriptsize $43.0 {}^{+1.6}_{-0.2} \pm 0.8$} 
& {\scriptsize $22.9 \pm 0.5 {}^{+0.7}_{-0.9}$}
& {\scriptsize $65.9 {}^{+2.1}_{-0.7} {}^{+1.5}_{-1.7}$}
\\
{\scriptsize 8 TeV}  & {\scriptsize $56.4 {}^{+2.1}_{-0.3} \pm 1.1$} 
& {\scriptsize $30.7 \pm 0.7 {}^{+0.9}_{-1.1}$}
& {\scriptsize $87.2 {}^{+2.8}_{-1.0} {}^{+2.0}_{-2.2}$}
\\ 
{\scriptsize 14 TeV} & {\scriptsize $154 {}^{+4}_{-1} \pm 3$} 
& {\scriptsize $94 {}^{+2}_{-1} {}^{+2}_{-3}$} 
& {\scriptsize $248 {}^{+6}_{-2} {}^{+5}_{-6}$} 
\\
\bottomrule
\end{tabular} 
\caption{$t$-channel cross sections in pb at the LHC for $m_t=173$ GeV.}
\label{t-channel}
\end{table}

The $t$-channel cross sections at LHC energies for single top production, 
single antitop production, and their sum are given in Table 1.
The $t$-channel single top quark production at the Tevatron is 
$1.04 {}^{+0.00}_{-0.02} \pm 0.06$ pb; the result for antitop 
at the Tevatron is the same.

\begin{figure}[htb]
  \centering
  \includegraphics[width=0.38\textwidth]{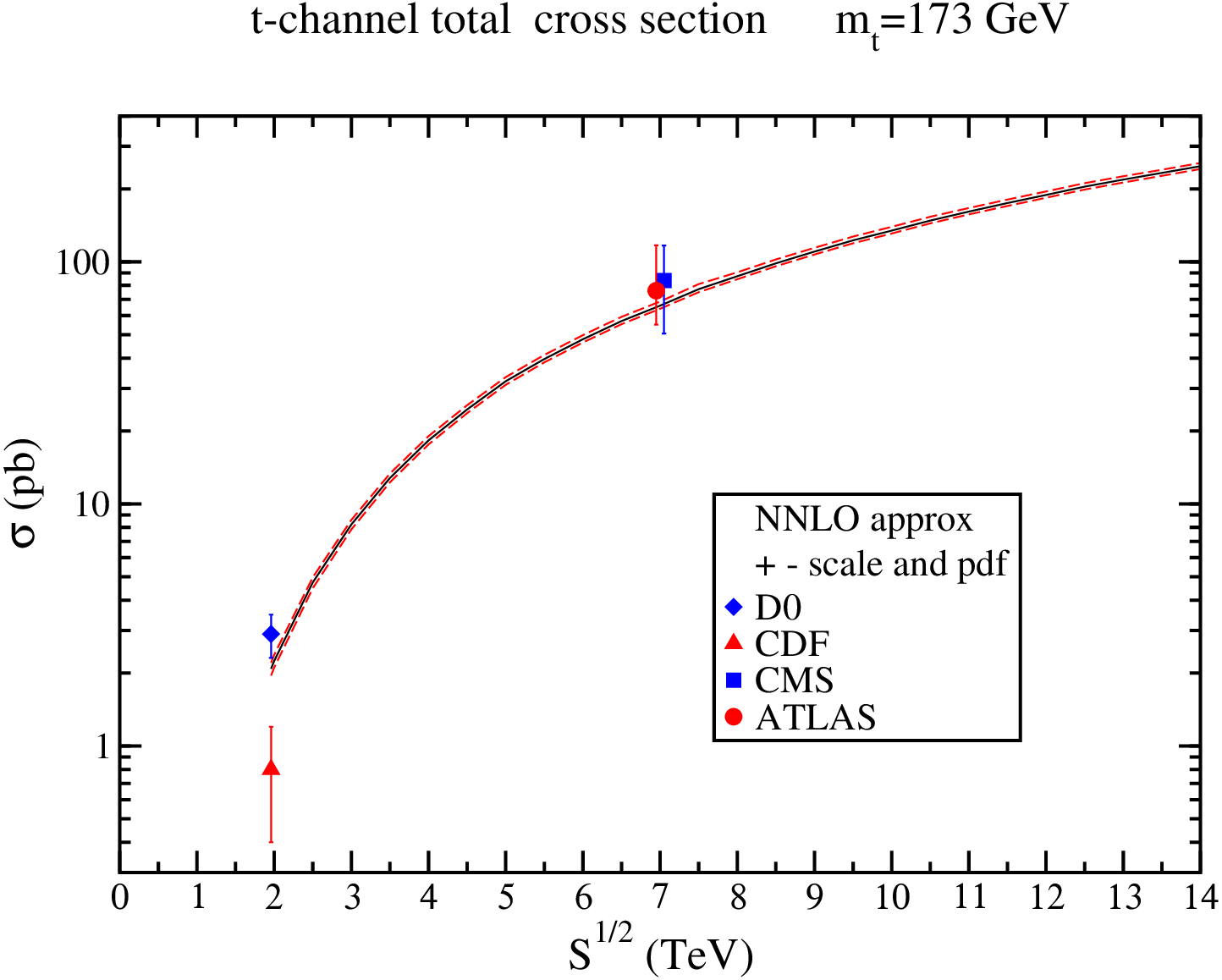}
\hspace{10mm}
  \includegraphics[width=0.38\textwidth]{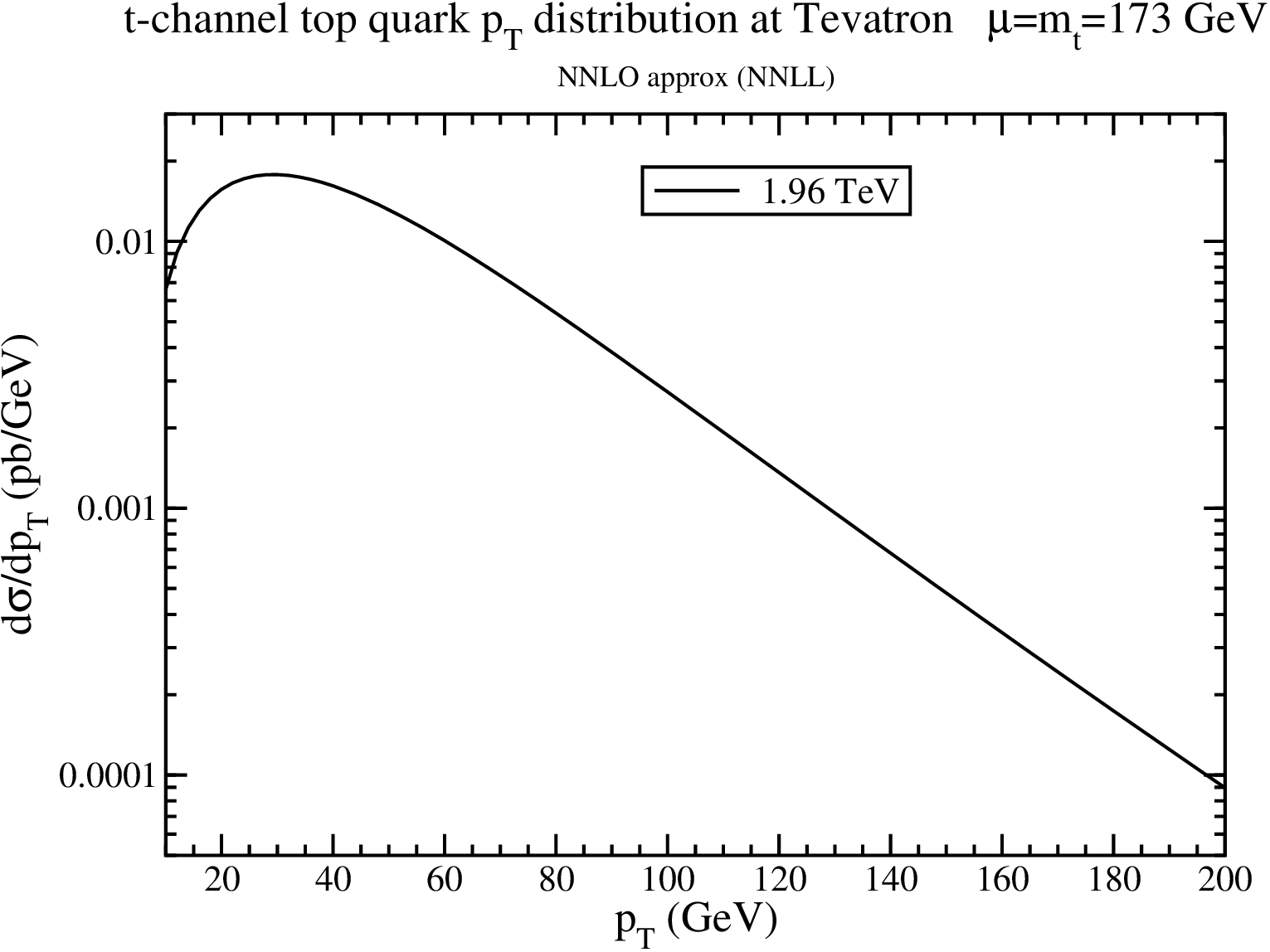}
  \caption{$t$-channel total cross section versus collider energy (left) and $t$-channel top quark $p_T$ distribution at the Tevatron (right).}
  \label{tcht}
\end{figure}

Results for the $t$-channel total cross section are shown versus collider energy in the 
left plot of Fig. 3. The right plot of Fig. 3 displays new results for the $t$-channel 
top quark $p_T$ distribution at the Tevatron. 

\begin{figure}[htb]
  \centering
  \includegraphics[width=0.38\textwidth]{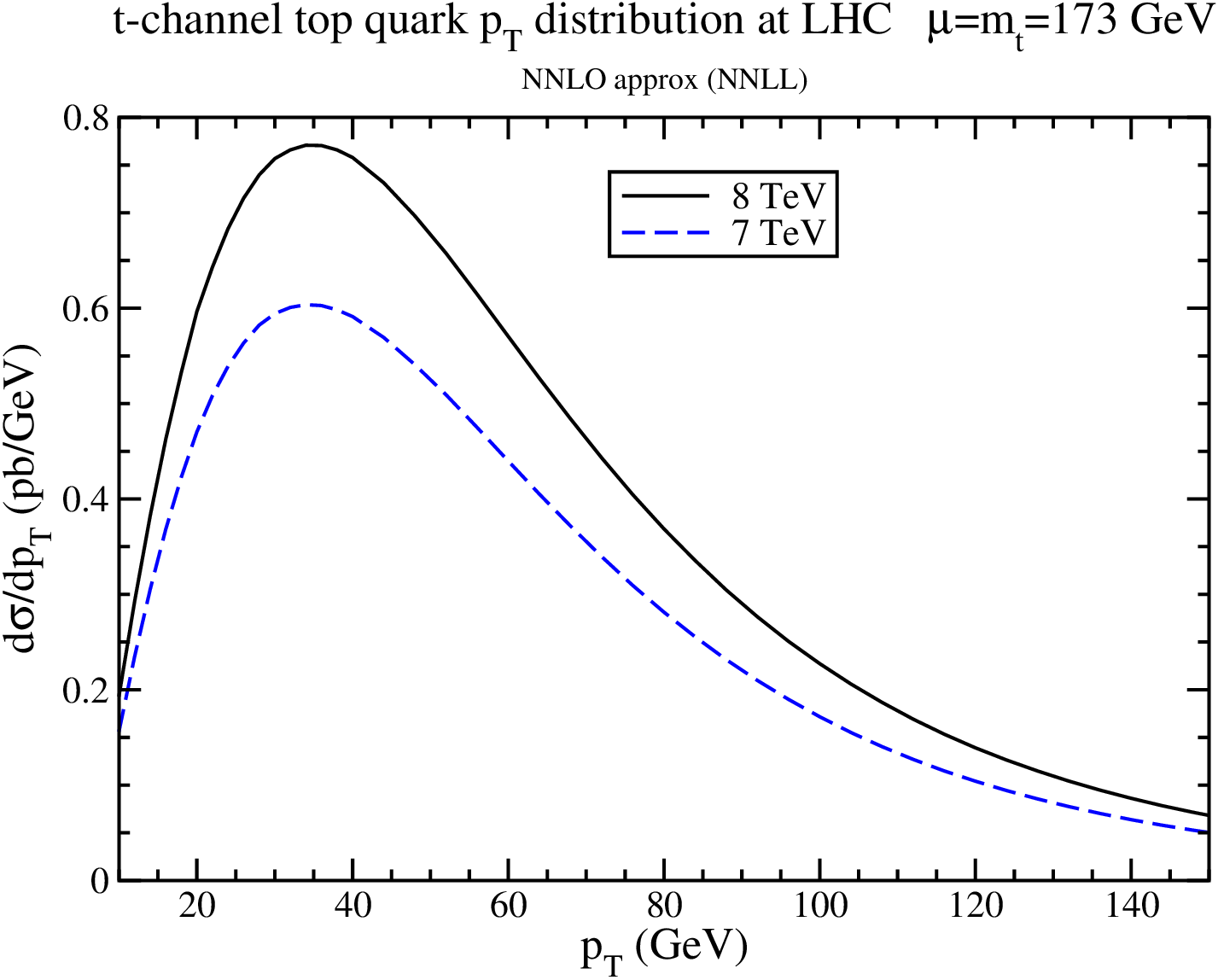}
\hspace{10mm}
  \includegraphics[width=0.38\textwidth]{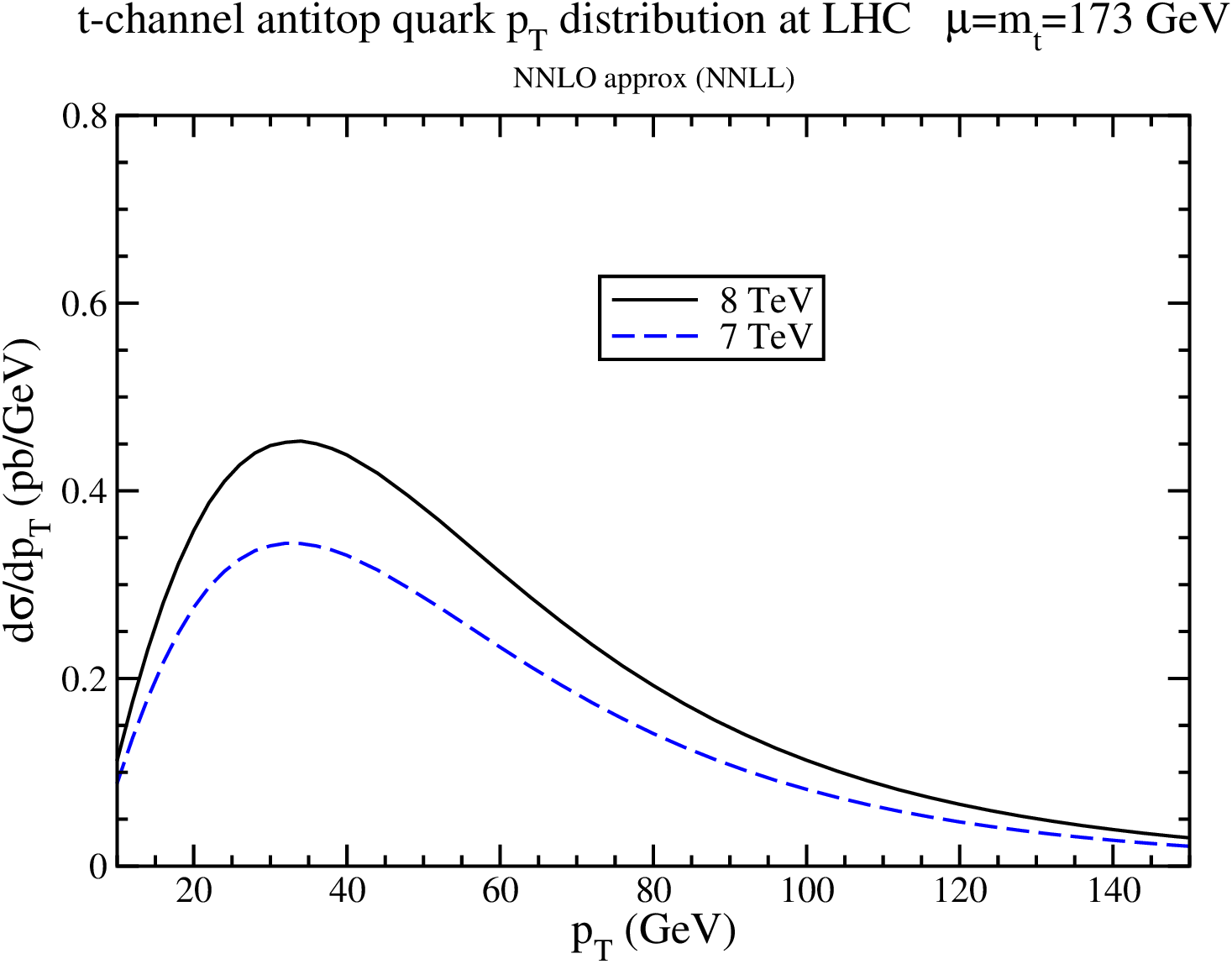}
  \caption{$t$-channel top (left) and antitop (right) $p_T$ distributions at the LHC.}
  \label{tchtoppTlhc}
\end{figure}

New results for $t$-channel top and antitop $p_T$ distributions at the LHC
are shown in Fig. 4. 

\begin{table}
  \centering
\begin{tabular}{c|c|c|c}
\toprule
{\scriptsize $s$-channel LHC}  & {\scriptsize $t$} &   
{\scriptsize ${\bar t}$} & {\scriptsize Total} \\ 
\midrule
{\scriptsize 7 TeV}  & {\scriptsize $3.14 \pm 0.06 {}^{+0.12}_{-0.10}$} 
& {\scriptsize $1.42 \pm 0.01 {}^{+0.06}_{-0.07}$}
& {\scriptsize $4.56 \pm 0.07 {}^{+0.18}_{-0.17}$}
\\
{\scriptsize 8 TeV}  & {\scriptsize $3.79 \pm 0.07 \pm 0.13$} 
& {\scriptsize $1.76 \pm 0.01 \pm 0.08$}
& {\scriptsize $5.55 \pm 0.08 \pm 0.21$}
\\ 
{\scriptsize 14 TeV} & {\scriptsize $7.87 \pm 0.14 {}^{+0.31}_{-0.28}$} 
& {\scriptsize $3.99 \pm 0.05 {}^{+0.14}_{-0.21}$}
& {\scriptsize $11.86 \pm 0.19 {}^{+0.45}_{-0.49}$} 
\\
\bottomrule
\end{tabular} 
\caption{$s$-channel cross sections in pb at the LHC for $m_t=173$ GeV.}
\label{s-channel}
\end{table}

Next we present $s$-channel results. Table 2 shows the 
$s$-channel cross sections at the LHC.
The $s$-channel total cross section versus LHC energy in shown in the left 
plot of Fig. 5.
The $s$-channel single top cross section at the Tevatron is $0.523{}^{+0.001}_{-0.005}{}^{+0.030}_{-0.028}$ pb; the result for antitop production 
at the Tevatron is identical to that for top.

\begin{figure}[htb]
  \centering
  \includegraphics[width=0.38\textwidth]{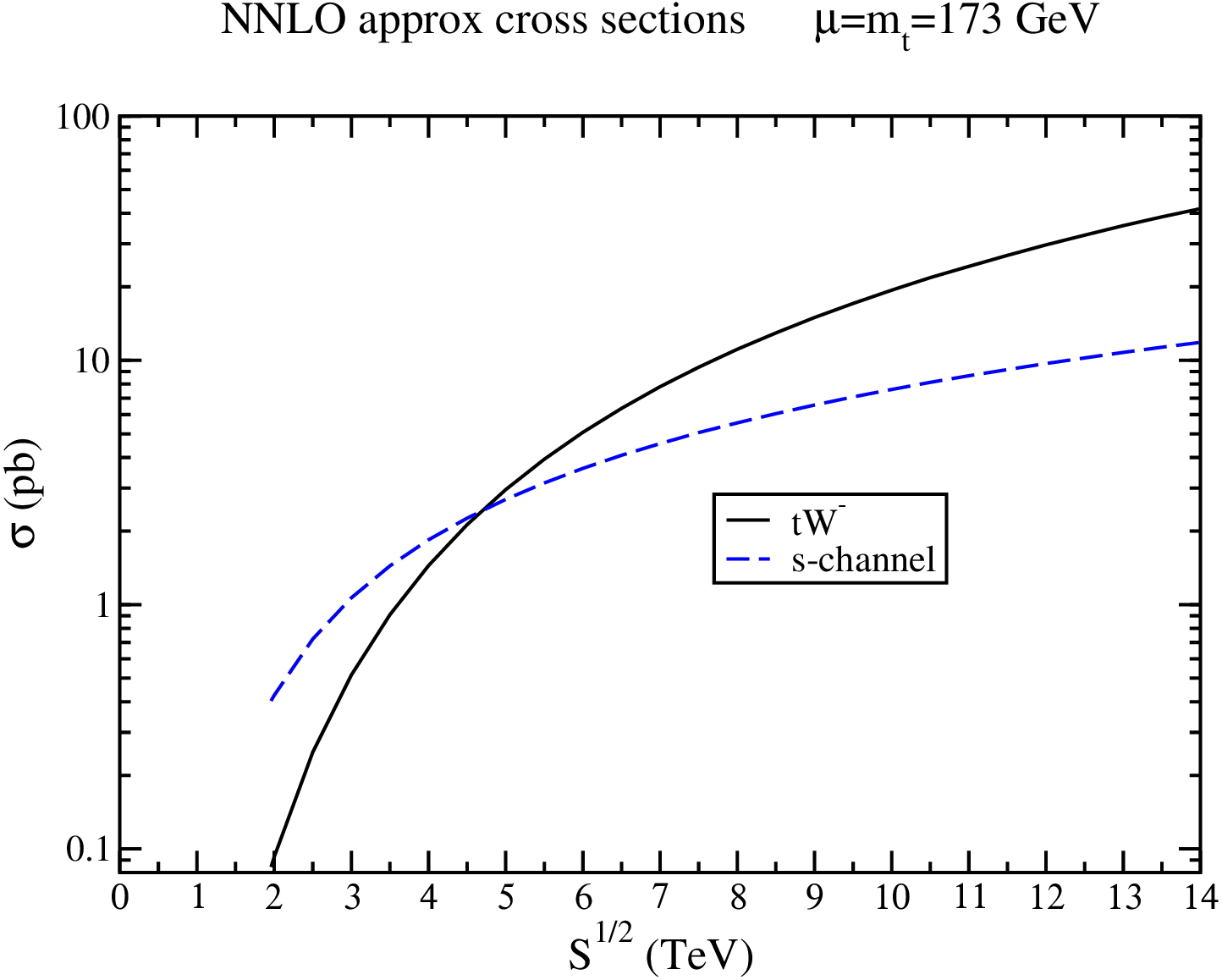}
\hspace{10mm}
  \includegraphics[width=0.38\textwidth]{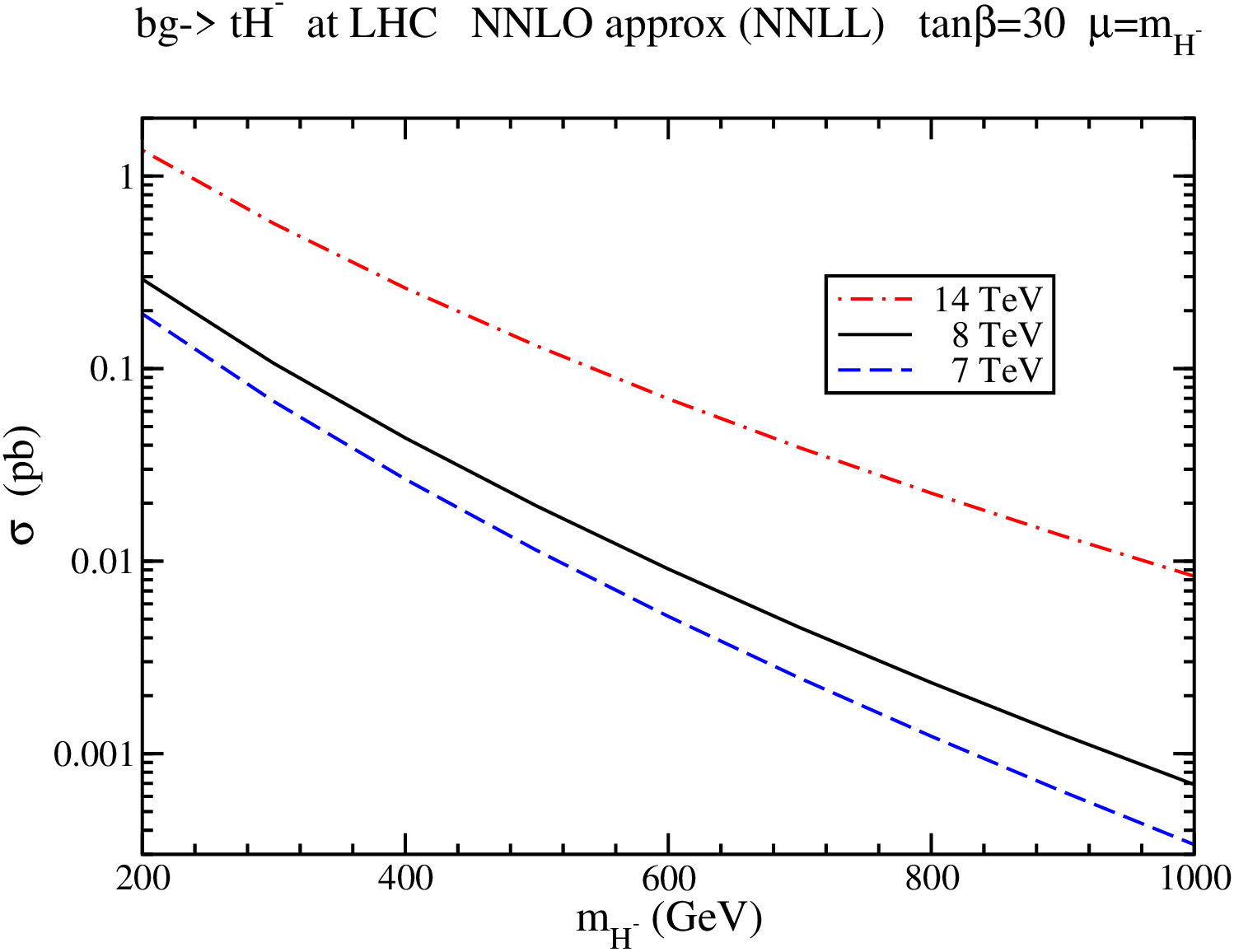}
  \caption{$s$-channel total and $tW^-$ (left) and $tH^-$ (right) production cross sections.}
  \label{tW}
\end{figure}

Next we study the associated $tW^-$ production at the LHC.
The $tW^-$ cross section at 7 TeV is $7.8 \pm 0.2 {}^{+0.5}_{-0.6}$ pb; 
at 8 TeV it is $11.1 \pm 0.3 \pm 0.7$ pb; 
and at 14 TeV it is $41.8 \pm 1.0 {}^{+1.5}_{-2.4}$ pb.
The cross section for ${\bar t}W^+$ production is identical.
The $tW^-$ cross section versus LHC energy in shown in the left 
plot of Fig. 5.

Finally, we study the associated production of a top quark with a charged 
Higgs. The right plot of Fig. 5 shows results at LHC energies 
versus charged Higgs mass with $\tan\beta=30$.

\section*{Acknowledgements}
The work of N.K. was supported by the National Science Foundation under
Grant No. PHY 0855421.

{\raggedright
\begin{footnotesize}

\end{footnotesize}
}

\end{document}